\documentclass[letterpaper]{llncs} 

\usepackage{fancyhdr}

\usepackage{color}
\usepackage[cmex10]{amsmath} 
\usepackage{amssymb}
\usepackage{bbm} 
\usepackage{graphicx} 
\usepackage[font=footnotesize,labelfont=bf]{caption} 
\usepackage[caption=false,font=footnotesize]{subfig} 
\usepackage[rightcaption]{sidecap} 
\usepackage{floatrow}
\usepackage{tabularx}
\usepackage{fixltx2e} 
\usepackage{engord} 
\usepackage{datetime} 
\usepackage{enumitem}
\usepackage{xifthen} 
\usepackage{multicol} 

\usepackage[linesnumbered,ruled]{algorithm2e}
\SetAlFnt{\scriptsize} 
\SetKwBlock{KwFunction}{function}{}
\SetKwBlock{KwOn}{on}{}
\SetKwBlock{KwPeriodically}{periodically}{}
\SetKwBlock{KwState}{state}{}
\SetKwBlock{KwAtomic}{atomically}{}
\SetKwBlock{KwConcurrently}{concurrently}{}
\SetCommentSty{textup} 
\SetKwComment{KwIttayComment}{(}{)} 
\SetKw{KwAnd}{and}
\SetKw{KwOr}{or}
\SetKw{KwSetTimer}{setTimer}
\SetKwBlock{KwExpTimer}{on timer expiration}{}
\SetKwFor{ForAll}{forall}{}

\ifx\pdftexversion\undefined
\usepackage[colorlinks,linkcolor=black,filecolor=black,citecolor=black,urlcolor=black,pdfstartview=FitH]{hyperref}
\else
\usepackage[colorlinks,linkcolor=blue,filecolor=blue,citecolor=blue,urlcolor=blue,pdfstartview=FitH]{hyperref}
\fi

\newtheorem{observation}{Observation}

\newcommand{\negspace}{\vspace{-0.5\baselineskip}}

\renewenvironment{thebibliography}[1]{%
  \begin{oldthebibliography}{#1}%
    \negspace
}%
{%
  \end{oldthebibliography}%
}

\title{ 
Majority is not Enough: \\
Bitcoin Mining is Vulnerable
} 

\date{}

    \author{Ittay Eyal \and Emin G\"un Sirer} 
    \institute{Department of Computer Science, Cornell University \\
               ittay.eyal@cornell.edu, egs@systems.cs.cornell.edu \\
               \negspace\negspace\negspace} 
           
\begin{document} 

\maketitle


\begin{abstract} 
The Bitcoin cryptocurrency records its transactions in a public log called the blockchain. Its security rests critically on the distributed protocol that maintains the blockchain, run by participants called miners. 
Conventional wisdom asserts that the protocol is incentive-compatible and secure against colluding minority groups, i.e., it incentivizes miners to follow the protocol as prescribed. 

We show that the Bitcoin protocol is not incentive-compatible. We present an attack with which colluding miners obtain a revenue larger than their fair share. 
This attack can have significant consequences for Bitcoin: Rational miners will prefer to join the selfish miners, and the colluding group will increase in size until it becomes a majority. At this point, the Bitcoin system ceases to be a decentralized currency. 

Selfish mining is feasible for any group size of colluding miners.
We propose a practical modification to the Bitcoin protocol that protects against selfish mining pools that command less than $1/4$ of the resources. 
This threshold is lower than the wrongly assumed $1/2$ bound, but better than the current reality where a group of any size can compromise the system. 
\end{abstract} 


    \section{Introduction} \label{sec:intro}
    
Bitcoin~\cite{nakamoto2008bitcoin} is a cryptocurrency that has recently emerged as a popular medium of exchange, with a rich and extensive ecosystem. 
The Bitcoin network runs at over $42\times 10^{18}$ FLOPS~\cite{bitcoinCharts2013stats}, with a total market capitalization around~1.5 billion US Dollars as of October~2013~\cite{blockchain2013marketCap}. 
Central to Bitcoin's operation is a global, public log, called the {\em blockchain}, that records all transactions between Bitcoin clients.
The security of the blockchain is established by a chain of cryptographic puzzles, solved by a loosely-organized network of participants called {\em miners}. 
Each miner that successfully solves a cryptopuzzle is allowed to record a set of transactions, and to collect a reward in Bitcoins.
The more \emph{mining power} (resources) a miner applies, the better are its chances to solve the puzzle first. 
This reward structure provides an incentive for miners to contribute their resources to the system, and is essential to the currency's decentralized nature. 

The Bitcoin protocol requires a majority of the miners to be \emph{honest}; that is, follow the Bitcoin protocol as prescribed. 
By construction, if a set of colluding miners comes to command a majority of the mining power in the network, the currency stops being decentralized and becomes
controlled by the colluding group.  Such a group can, for example, prohibit certain transactions, or all of them. 
It is, therefore, critical that the protocol be designed such that miners have no incentive to form such large colluding groups.

Empirical evidence shows that Bitcoin miners behave strategically and form pools. 
Specifically, because rewards are distributed at infrequent, random intervals, miners form mining pools in order to decrease the variance of their income rate. 
Within such pools, all members contribute to the solution of each cryptopuzzle, and share the rewards proportionally to their contributions. 
To the best of our knowledge, so far such pools have been benign and followed the protocol. 
    
Indeed, conventional wisdom has long asserted that the Bitcoin protocol is incentive-compatible; that is, the best strategy of a rational minority pool is to be honest, and a minority of colluding miners cannot earn disproportionate benefits by deviating from the protocol~\cite{barber2012bitter}. 
Because the protocol is believed to reward miners strictly in proportion to the ratio of the overall mining power they control, 
a miner in a large pool is believed to earn the same revenue as it would in a small pool. Consequently, 
there is no advantage for colluding miners to organize into ever-increasing pools. 
Therefore, pool formation by honest rational miners poses no threat to the system. 

In this paper, we show that the conventional wisdom is wrong: the Bitcoin protocol, as prescribed and implemented, is not incentive-compatible.
We describe a strategy that can be used by a minority pool to obtain more revenue than the pool's fair share, that is, more than its ratio of the total mining power. 

The key idea behind this strategy, called Selfish Mining, is for a pool to keep its discovered blocks private, thereby intentionally forking the chain. The honest nodes
continue to mine on the public chain, while the pool mines on its own private branch. If the pool discovers more blocks, it develops a longer lead on the public chain,
and continues to keep these new blocks private. When the public branch approaches the pool's private branch in length, the selfish miners reveal blocks from their private chain to the public. 

This strategy leads honest miners that follow the Bitcoin protocol to waste resources on mining cryptopuzzles that end up serving no purpose. 
Our analysis demonstrates that, while both honest and selfish parties waste some resources, the honest miners waste proportionally more, and the selfish pool's rewards exceed its share of the network's mining power, conferring it a competitive advantage and incentivizing rational miners to join the selfish mining pool.

We show that, above a certain threshold size, the revenue of a selfish pool rises superlinearly with pool size above its revenue with the honest strategy. The implications of this statement are devastating for the system. Once a selfish mining pool reaches the threshold, rational miners will preferentially join selfish miners to reap the higher revenues compared to other pools. Such a selfish mining pool will quickly grow to become a majority, at which point the pool will be the only creator of blocks, the decentralized nature of the currency will collapse, and a single entity, the selfish pool manager, will control the system. 

Since a selfish mining pool that exceeds threshold size poses a threat to the Bitcoin system, we characterize how the threshold varies as a function of 
message propagation speed in the network. We show that, for a mining pool with high connectivity and good control on information flow, the threshold is close to zero. This implies
that the Bitcoin system is safe only when $100$\% of the miners are honest. The first selfish miner will earn proportionally higher revenues than its honest counterparts, and the revenue of the selfish mining pool will increase superlinearly with pool size. 

We further show that the upper bound on threshold size is~$1/3$: the protocol will never be safe against attacks by a selfish mining pool that commands more than $33$\% of the total mining power of the network. This upper bound is substantially lower than the $50$\% figure currently assumed, and difficult to achieve in practice.
Finally, we suggest a simple modification to the Bitcoin protocol that achieves a threshold of~$1/4$. This change is backwards-compatible and \emph{progressive}; that is, it can be adopted by current clients with modest changes, does not require full adoption to provide a benefit and partial adoption will proportionally increase the threshold. 

In summary, the contributions of this work are: 
\begin{enumerate} 
\item Introduction of the Selfish-Mine strategy, which demonstrates that Bitcoin mining is not incentive compatible (Section~\ref{sec:algo}). 

\item Analysis of Selfish-Mine, and when it can benefit a pool (Section~\ref{sec:analysis}). 

\item Analysis of majority-pool formation in face of selfish mining (Section~\ref{sec:poolFormation}). 

\item A simple backward-compatible progressive modification to the Bitcoin protocol that would raise the threshold from zero to~$1/4$ (Section~\ref{sec:btcPatch}). 
\end{enumerate} 

We are unaware of previous work that addresses the security of the blockchain. We provide an overview of related work in Section~\ref{sec:related}, and discuss the implications of our results in Section~\ref{sec:discussion}. 


    \section{Preliminaries} \label{sec:prelim} 
\negspace

Bitcoin is a distributed, decentralized crypto-currency~\cite{bitcoin2013protocol,bitcoin2013rules,nakamoto2008bitcoin,bitcoin2013source}. 
The users of Bitcoin are called \emph{clients}, each of whom can command accounts, known as \emph{addresses}. A client can send Bitcoins to another client by forming a transaction and committing it into a global append-only log called the \emph{blockchain}. The blockchain is maintained by a network of \emph{miners}, which are compensated for their effort in Bitcoins. Bitcoin transactions are protected with cryptographic techniques that ensure only the rightful owner of a Bitcoin address can transfer funds from it. 

The miners are in charge of recording the transactions in the blockchain, which determines the ownership of Bitcoins. 
A client owns~$x$ Bitcoins at time~$t$ if, in the prefix of the blockchain up to time~$t$, the aggregate of transactions involving that client's address amounts to~$x$. 
Miners only accept transactions if the balance at the source is sufficient. 

        \subsection{Blockchain and Mining} 
\negspace

The blockchain records the transactions in units of blocks. Each block includes a unique ID, and the ID of the preceding block. The first block, dubbed \emph{the genesis block}, is defined as part of the protocol. A valid block contains a solution to a cryptopuzzle involving the hash of the previous block, the hash of the transactions in the current block, and a Bitcoin address which is to be credited with a reward for solving the cryptopuzzle. This process is called Bitcoin \emph{mining}, and, by slight abuse of terminology, we refer to the creation of blocks as \emph{block mining}. 
The specific cryptopuzzle is a double-hash whose result has to be smaller than a set value. The problem difficulty, set by this value, is dynamically adjusted such that blocks are generated at an average rate of one every ten minutes. 

Any miner may add a valid block to the chain by simply publishing it over an overlay network to all other miners. 
If two miners create two blocks with the same preceding block, the chain is \emph{forked} into two \emph{branches}, forming a tree. Other miners may subsequently add new valid blocks to either branch. When a miner tries to add a new block after an existing block, we say it \emph{mines on} the existing block. This existing block may be the head of a branch, in which case we say the miner mines on the head of the branch, or simply on the branch. 

The formation of branches is undesirable since the miners have to maintain a globally-agreed totally ordered set of transactions. To resolve forks, the protocol prescribes miners to adopt and mine on the longest chain.\footnote{The criterion is actually the most difficult chain in the block tree, i.e., the one that required (in expectancy) the most mining power to create. To simplify presentation, and because it is usually the case, we assume the set difficulty at the different branches is the same, and so the longest chain is also the most difficult one.} All miners add blocks to the longest chain they know of, or the first one they heard of if there are branches of equal length. This causes forked branches to be pruned; transactions in pruned blocks are ignored, and may be resubmitted by clients.

We note that block dissemination over the overlay network takes seconds, whereas the average mining interval is~10 minutes. Accidental bifurcation is therefore rare, and occurs on average once about every~60 blocks~\cite{decker2013propagation}. 

When a miner creates a block, it is compensated for its efforts with Bitcoins. This compensation includes a per-transaction fee paid by the users whose transactions are included, as well as an amount of new Bitcoins that did not exist before.\footnote{The rate at which the new Bitcoins are generated is designed to slowly decrease towards zero, and will reach zero when almost~21 million Bitcoins are created. Then, the miners' revenue will be only from transaction fees. }

        \subsection{Pool formation} 
\negspace
        
The probability of mining a block is proportional to the computational resources used for solving the associated cryptopuzzle.
Due the nature of the mining process, the interval between mining events exhibits high variance from the point of view of a single miner. 
A single home miner using a dedicated ASIC is unlikely to mine a block for years~\cite{swanson2013calculator}. 
Consequently, miners typically organize themselves into mining \emph{pools}. 
All members of a pool work together to mine each block, and share their revenues when one of them successfully mines a block. 
While joining a pool does not change a miner's expected revenue, it decreases the variance and makes the monthly revenues more predictable. 


    \section{The Selfish-Mine Strategy} \label{sec:algo}

First, we formalize a model that captures the essentials of Bitcoin mining behavior and introduces notation for relevant system parameters. Then we detail the selfish mining algorithm. 

        \subsection{Modeling Miners and Pools} 

The system is comprised of a set of miners $1, \dots, n$. Each miner~$i$ has mining power~$m_i$, such that $\sum_{i=1}^n m_i = 1$. Each miner chooses a chain head to mine, and finds a subsequent block for that head after a time interval that is exponentially distributed with mean~$m_i^{-1}$. We assume that miners are rational; that is, they try to maximize their revenue, and may deviate from the protocol to do so. 

\newcommand{\poolSet}{\ensuremath{ P }}
\newcommand{\pool}{\ensuremath{ p }}
\newcommand{\poolNum}{\ensuremath{ l }}

A group of miners can form a pool that behaves as single agent with a centralized coordinator, following some strategy. The mining power of a pool is the sum of mining power of its members, and its revenue is divided among its members according to their relative mining power. 
The \emph{expected relative revenue}, or simply the \emph{revenue} of a pool is the expected fraction of blocks that were mined by that pool out of the total number of blocks in the longest chain. 

        \subsection{Selfish-Mine}

We now describe our strategy, called Selfish-Mine. As we show in Section~\ref{sec:analysis}, Selfish-Mine allows a pool of sufficient size to obtain a revenue larger than its ratio of mining power. 
    For simplicity, and without loss of generality, we assume that miners are divided into two groups, a colluding minority pool that follows the selfish mining strategy,
and a majority that follows the honest mining strategy (others). It is immaterial whether the honest miners operate as a single group, as a collection of groups, or individually.

The key insight behind the selfish mining strategy is to force the honest miners into performing wasted computations on the stale public branch. Specifically, selfish mining forces the honest miners to spend their cycles on blocks that are destined to not be part of the blockchain. 

Selfish miners achieve this goal by selectively revealing their mined blocks to invalidate the honest miners' work. 
Approximately speaking, the selfish mining pool keeps its mined blocks private, secretly bifurcating the blockchain and creating a private branch. 
Meanwhile, the honest miners continue mining on the shorter, public branch. 
Because the selfish miners command a relatively small portion of the total mining power, their private branch will not remain ahead of the public branch indefinitely. 
Consequently, selfish mining judiciously reveals blocks from the private branch to the public, such that the honest miners will switch to the recently revealed blocks,
abandoning the shorter public branch. 
This renders their previous effort spent on the shorter public branch wasted, and enables the selfish pool to collect higher revenues by incorporating a higher fraction
of its blocks into the blockchain. 

\newcommand{\privateBranchLen}{\textit{privateBranchLen}} 
\newcommand{\prevDelta}{\ensuremath{ \Delta_{\textit{prev}} }} 

\begin{algorithm}[t] 
\SetAlgoNoLine 
\SetAlgoNoEnd 
\DontPrintSemicolon 
\caption{Selfish-Mine} 
\label{alg:btcProc} 
\KwOn(Init) { 
    public chain $\gets$ publicly known blocks \; 
    private chain $\gets$ publicly known blocks \; 
    $\privateBranchLen \gets 0$ \; 
    Mine at the head of the private chain. 
} 
\BlankLine \BlankLine 

\KwOn(My pool found a block) {
    $\prevDelta \gets \text{length}(\text{private chain}) - \text{length}(\text{public chain})$ \; 
    append new block to private chain \; 
    $\privateBranchLen \gets \privateBranchLen + 1$ \; 
    \If({\hfill(Was tie with branch of 1)}){$\prevDelta = 0$ and $\privateBranchLen = 2$} { 
        publish all of the private chain \KwIttayComment*{Pool wins due to the lead of 1} 
        $\privateBranchLen \gets 0$ \; 
    }
    Mine at the new head of the private chain. 
} 
\BlankLine \BlankLine 

\KwOn(Others found a block) { 
    $\prevDelta \gets \text{length}(\text{private chain}) - \text{length}(\text{public chain})$ \; 
    append new block to public chain \; 
    \If{$\prevDelta = 0$} { 
        private chain $\gets$ public chain \KwIttayComment*{they win} 
        $\privateBranchLen \gets 0$ \; 
    } \ElseIf{$\prevDelta = 1$} {
        publish last block of the private chain \KwIttayComment*{Now same length. Try our luck} 
    } \ElseIf{$\prevDelta = 2$} {
        publish all of the private chain \KwIttayComment*{Pool wins due to the lead of 1} 
        $\privateBranchLen \gets 0$ \; 
    } \Else({\hfill($\prevDelta > 2$)}) { 
        publish first unpublished block in private block. 
    }
    Mine at the head of the private chain. 
} 
\end{algorithm} 

Armed with this intuition, we can fully specify the selfish mining strategy, shown in Algorithm~\ref{alg:btcProc}. The strategy is driven by mining events by 
the selfish pool or by the others. Its decisions depend only on the relative lengths of the selfish pool's private branch versus the public branch.
It is best to illustrate the operation of the selfish mining strategy by going through sample scenarios involving different public and private chain lengths.

When the public branch is longer than the private branch, the selfish mining pool is behind the public branch. Because of the power differential between the selfish miners 
and the others, the chances of the selfish miners mining on their own private branch and overtaking the main branch are small. Consequently, the selfish miner pool simply adopts
the main branch whenever its private branch falls behind. As others find new blocks and publish them, the pool updates and mines at the current public head. 

When the selfish miner pool finds a block, it is in an advantageous position with a single block lead on the public branch on which the honest miners operate. Instead of naively publishing this private block and notifying the rest of the miners of the newly discovered block, selfish miners keep this block private to the pool. There are two outcomes possible at this point: either the honest miners discover a new block on the public branch, nullifying the pool's lead, or else the pool mines a second block and extends its lead on the honest miners. 

In the first scenario where the honest nodes succeed in finding a block on the public branch, nullifying the selfish pool's lead, the pool immediately publishes its private branch (of length~1). This yields a toss-up where either branch may win. The selfish miners unanimously adopt and extend the previously private branch, while the honest miners will choose to mine on either branch, depending on the propagation of the notifications. If the selfish pool manages to mine a subsequent block ahead of the honest miners that did not adopt the pool's recently revealed block, it publishes immediately to enjoy the revenue of both the first and the second blocks of its branch. 
If the honest miners mine a block after the pool's revealed block, the pool enjoys the revenue of its block, while the others get the revenue from their block. Finally, if the honest miners mine a block after their own block, they enjoy the revenue of their two blocks while the pool gets nothing. 

In the second scenario, where the selfish pool succeeds in finding a second block, it develops a comfortable lead of two blocks that allow it with some cushion against discoveries by the honest miners. 
Once the pool reaches this point, it continues to mine at the head of its private branch. It publishes one block from its private branch for every block the others find. Since the selfish pool is a minority, its lead will eventually reduce to a single block with high probability. 
At this point, the honest miners are too close, so the pool publishes its private branch. Since the private branch is longer than the public branch by one block, it is adopted by all miners as the main branch, and the pool enjoys the revenue of all its blocks. This brings the system back to a state where there is just a single branch until the pool bifurcates it again. 


    \section{Analysis} \label{sec:analysis} 

We can now analyze the expected rewards for a system where the selfish pool has mining power of~$\alpha$ and the others of~$(1-\alpha)$. 

\begin{figure}[!t] 
\centering
\includegraphics[width=0.7\linewidth]{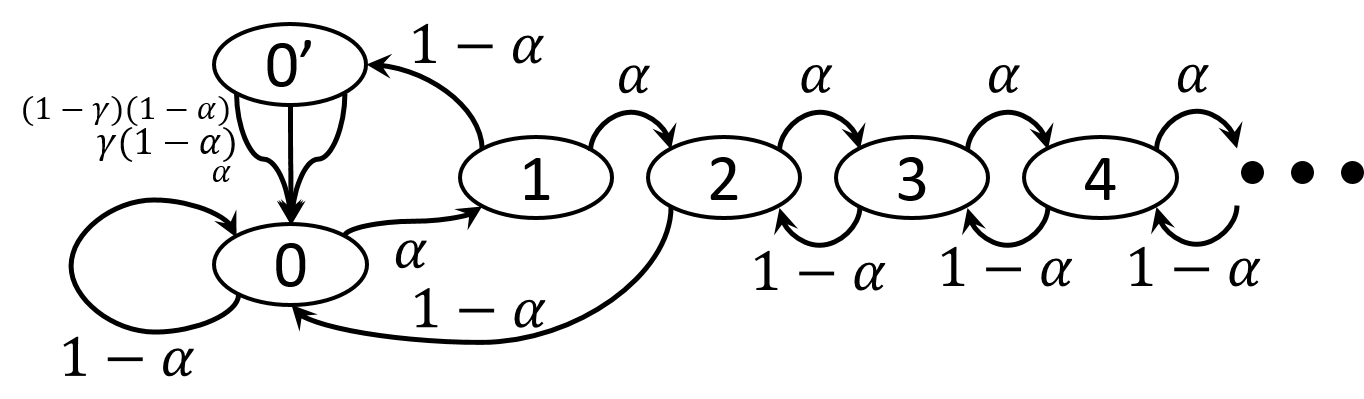}
\caption{ 
State machine with transition frequencies. 
} 
\label{fig:stateMachine} 
\end{figure}

Figure~\ref{fig:stateMachine} illustrates the progress of the system as a state machine. The states of the system represent the lead of the selfish pool; that is, the difference between the number of unpublished blocks in the pool's private branch and the length of the public branch. Zero lead is separated to states~0 and~0'. State~0 is the state where there are no branches; that is, there is only a single, global, public longest chain. State~0' is the state where there are two public branches of length one: the main branch, and the branch that was private to the selfish miners, and published to match the main branch. The transitions in the figure correspond to mining events, either by the selfish pool or by the others. Recall that these events occur at exponential intervals with an average frequency of $\alpha$ and $(1-\alpha)$, respectively. 

We can analyze the expected rewards from selfish mining by taking into account the frequencies associated with each state transition of the state machine, and calculating the corresponding rewards. Let us go through the various cases and describe the associated events that trigger state transitions. 

If the pool has a private branch of length~1 and the others mine one block, the pool publishes its branch immediately, which results in two public branches of length~1. Miners in the selfish pool all mine on the pool's branch, because a subsequent block discovery on this branch will yield a reward for the pool. The honest miners, following the standard Bitcoin protocol implementation, mine on the branch they heard of first. We denote by $\gamma$ the ratio of honest miners that choose to mine on the pool's block, and the other $(1-\gamma)$ of the non-pool miners mine on the other branch. 

For state $s = 0, 1, 2, \dots$, with frequency $\alpha$, the pool mines a block and the lead increases by one to $s+1$. In states $s = 3, 4, \dots$, with frequency $(1 - \alpha)$, the honest miners mine a block and the lead decreases by one to $s-1$. 
If the others mine a block when the lead is two, the pool publishes its private branch, and the system drops to a lead of~0. 
If the others mine a block with the lead is~1, we arrive at the aforementioned state~0'. From 0', there are three possible transitions, all leading to state~0 with total frequency~1: (1) the pool mines a block on its previously private branch (frequency $\alpha$), (2) the others mine a block on the previously private branch (frequency $\gamma(1-\alpha)$), and (3) the others mine a block on the public branch (frequency $(1-\gamma)(1-\alpha)$). 

        \subsection{State Probabilities}

We analyze this state machine to calculate its probability distribution over the state space. We obtain the following equations: 
\begin{equation} 
\left\{ \begin{array}{l}
\alpha p_0 = (1-\alpha) p_1 + (1-\alpha) p_2 \\
p_{0'} = (1 - \alpha) p_1 \\
\alpha p_1 = (1-\alpha) p_2 \\
\forall k \ge 2: \alpha p_k = (1-\alpha) p_{k+1} \\
\sum_{k=0}^\infty p_k + p_{0'} = 1 
\end{array} \right.
\label{eqn:stateEqns} 
\end{equation} 
Solving (\ref{eqn:stateEqns}) (See Appendix~\ref{app:probs} for details), we get: 
\begin{eqnarray}
&& p_0 = \frac{\alpha-2 \alpha^2}{\alpha(2\alpha^3 - 4\alpha^2 + 1)} \label{eqn:pFirst} \\
&& p_{0'} = \frac{(1 - \alpha)(\alpha - 2 \alpha^2)}{1 - 4\alpha^2 + 2\alpha^3} \\
&& p_1 = \frac{\alpha-2 \alpha^2}{2\alpha^3 - 4\alpha^2 + 1} \\
&& \forall k \ge 2: p_k =  \left( \frac{\alpha}{1-\alpha} \right)^{k-1} \frac{\alpha-2 \alpha^2}{2\alpha^3 - 4\alpha^2 + 1} \label{eqn:pLast}
\end{eqnarray} 

        \subsection{Revenue} 

The probability distribution over the state space provides the foundation for analyzing the revenue obtained by the selfish pool and by the honest miners. The revenue for finding a block belongs to its miner only if this block ends up in the main chain. We detail the revenues on each event below.

\begin{enumerate}[label=(\alph*)] 
\item \emph{Any state but two branches of length~1, pools finds a block.} 
The pool appends one block to its private branch, increasing its lead on the public branch by one. The revenue from this block will be determined later. \label{itm:pool} 

\centerline{
    \raisebox{-0.5\height}{\includegraphics[height=0.6in]{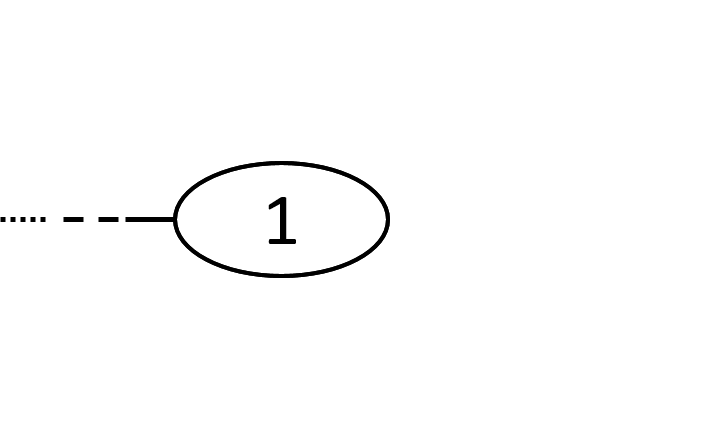}} 
    {\Large \hspace{0.0cm} $\Longrightarrow$ \hspace{0.0cm}}
    \raisebox{-0.5\height}{\includegraphics[height=0.6in]{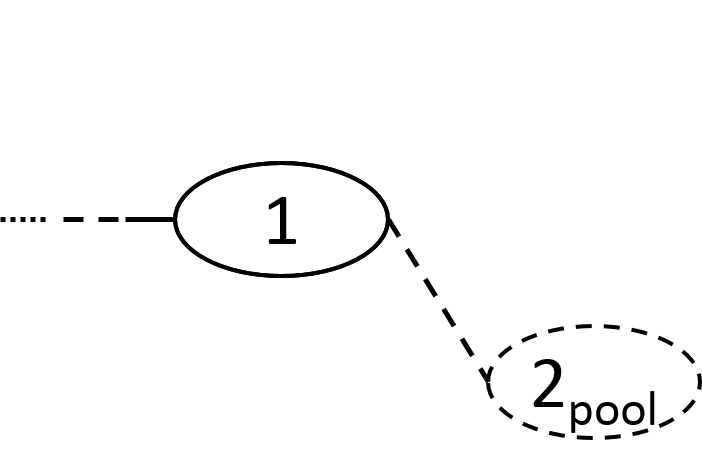}}
}

or

\centerline{
    \raisebox{-0.5\height}{\includegraphics[height=0.6in]{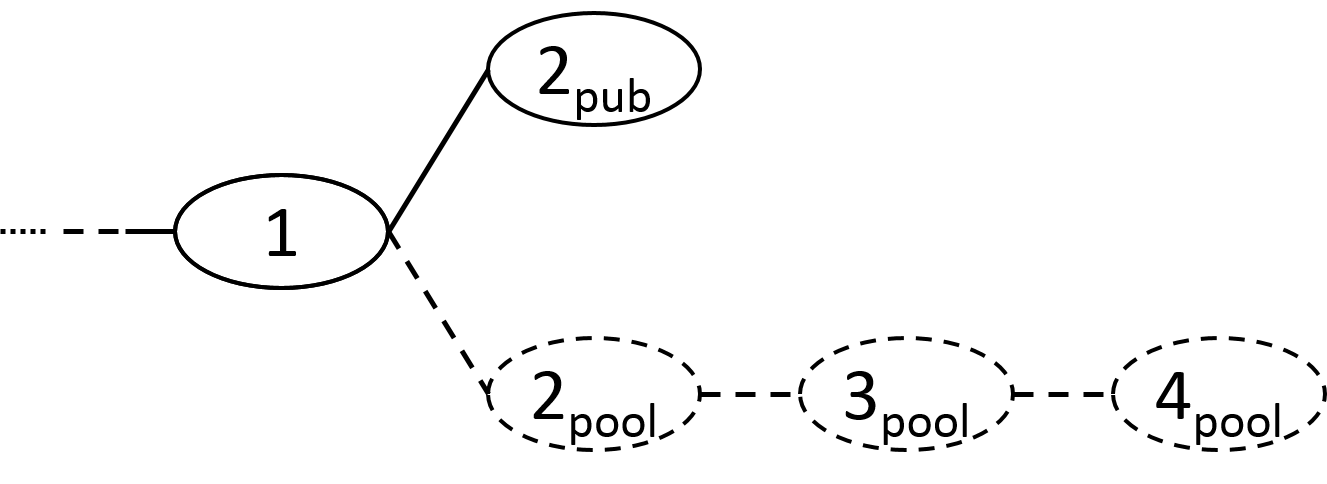}} 
    {\Large \hspace{0.0cm} $\Longrightarrow$ \hspace{0.0cm}}
    \raisebox{-0.5\height}{\includegraphics[height=0.6in]{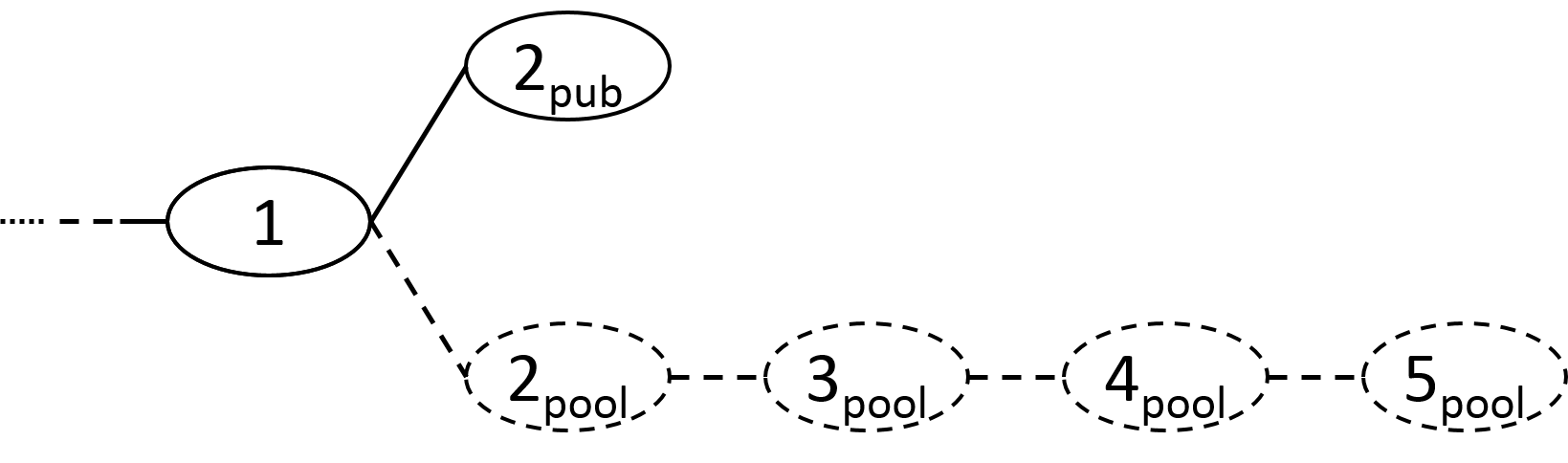}}
}

\vspace{0.5\baselineskip}

\item \emph{Was two branches of length~1, pools finds a block.}
The pool publishes its secret branch of length two, thus obtaining a revenue of two. \label{itm:poolPoolZeroPrime} 

\vspace{0.5\baselineskip}

\centerline{
    \raisebox{-0.5\height}{\includegraphics[height=0.6in]{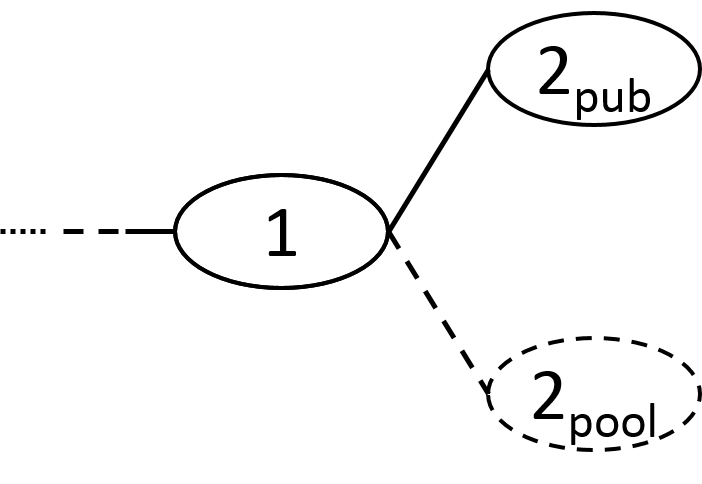}} 
    {\Large \hspace{0.0cm} $\Longrightarrow$ \hspace{0.0cm}}
    \raisebox{-0.5\height}{\includegraphics[height=0.6in]{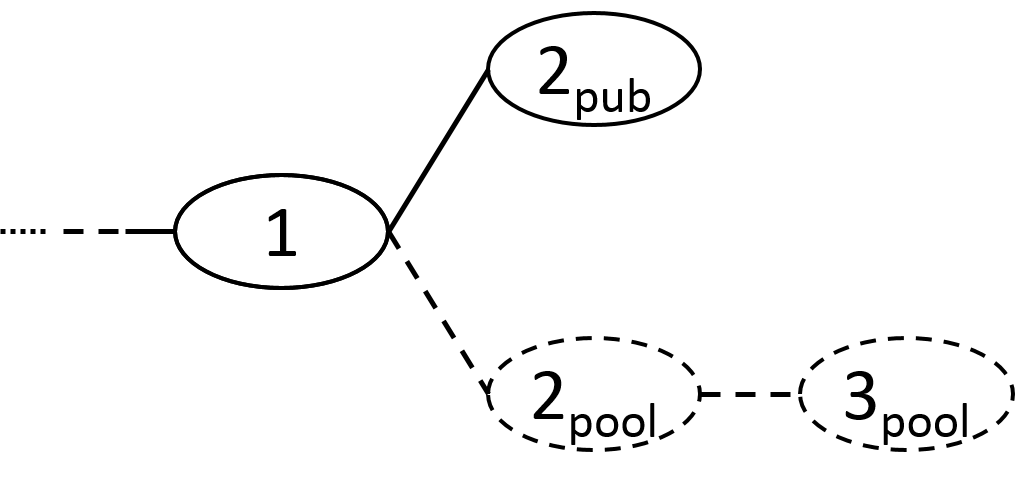}}
    {\Large \hspace{0.0cm} $\Longrightarrow$ \hspace{0.0cm}}
    \raisebox{-0.5\height}{\includegraphics[height=0.6in]{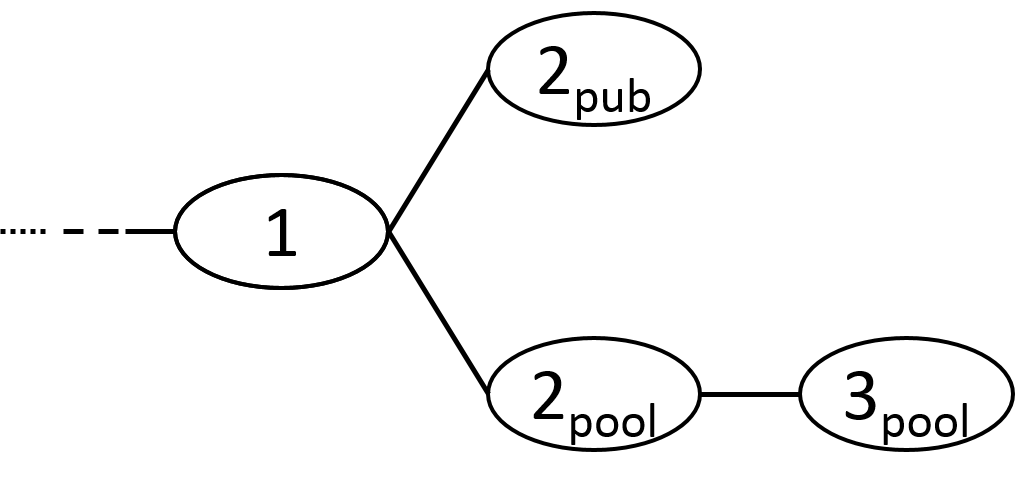}}
}

\vspace{0.5\baselineskip}

\item \emph{Was two branches of length~1, others find a block after pool head.} 
The pool and the others obtain a revenue of one each~--- the others for the new head, the pool for its predecessor. \label{itm:poolOthersZeroPrime}

\vspace{0.5\baselineskip}

\centerline{
    \raisebox{-0.5\height}{\includegraphics[height=0.6in]{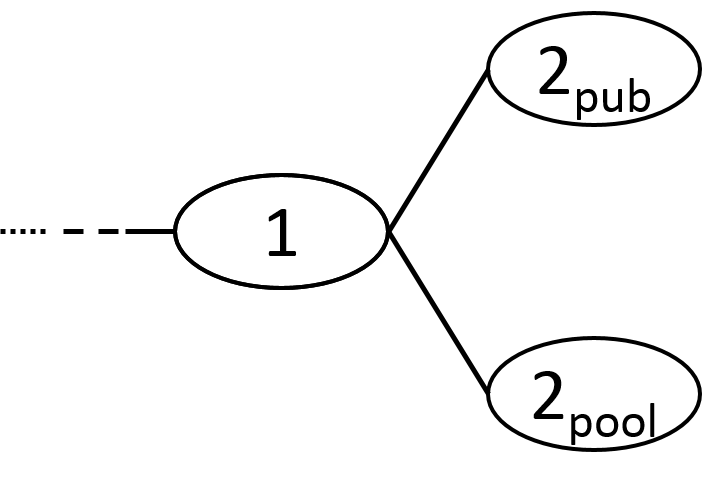}} 
    {\Large \hspace{0.0cm} $\Longrightarrow$ \hspace{0.0cm}}
    \raisebox{-0.5\height}{\includegraphics[height=0.6in]{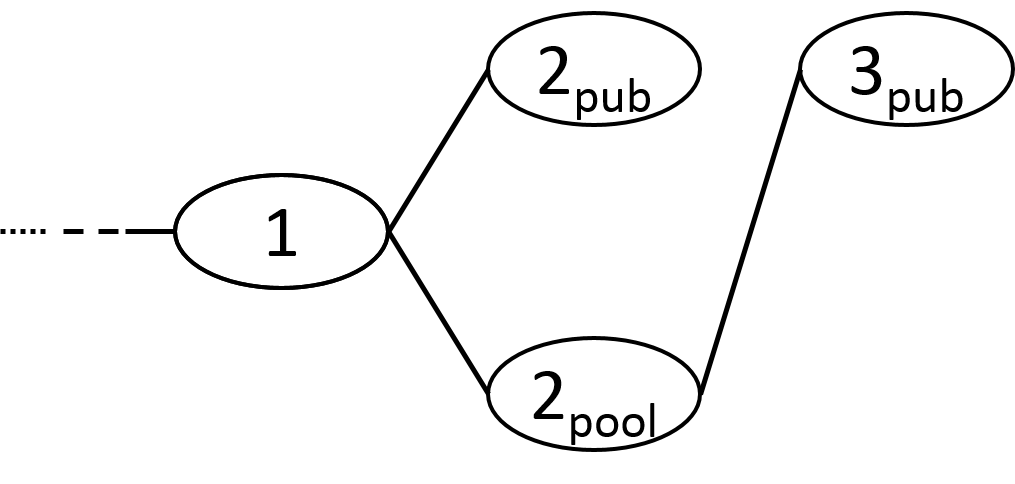}}
}

\vspace{0.5\baselineskip}

\item \emph{Was two branches of length~1, others find a block after others' head.} 
The others obtain a revenue of two. \label{itm:othersOthersZeroPrime}

\vspace{0.5\baselineskip}

\centerline{
    \raisebox{-0.5\height}{\includegraphics[height=0.6in]{branch1-1pub.png}} 
    {\Large \hspace{0.0cm} $\Longrightarrow$ \hspace{0.0cm}}
    \raisebox{-0.5\height}{\includegraphics[height=0.6in]{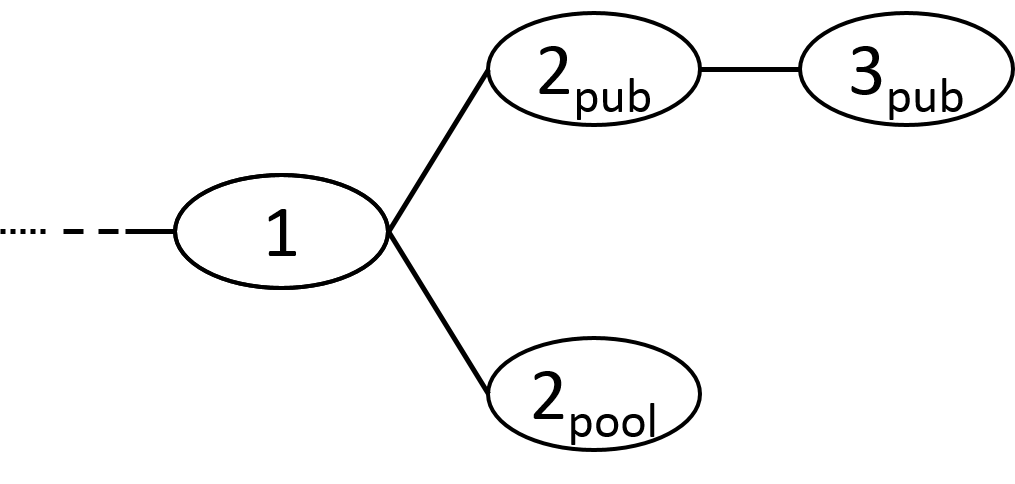}}
}

\vspace{0.5\baselineskip}

\item \emph{No private branch, others find a block.} The others obtain a revenue of one, and both the pool and the others start mining on the new head. \label{itm:others0}

\item \emph{Lead was~1, others find a block.} Now there are two branches of length one, and the pool publishes its single secret block. The pool tries to mine on its previously private head, and the others split between the two heads. Denote by $\gamma$ the ratio of others that choose the non-pool block. 

The revenue from this block cannot be determined yet, because it depends on which branch will win. It will be counted later. \label{itm:others1}

\vspace{0.5\baselineskip}

\centerline{
    \raisebox{-0.5\height}{\includegraphics[height=0.6in]{branch0-1.png}} 
    {\Large \hspace{0.0cm} $\Longrightarrow$ \hspace{0.0cm}}
    \raisebox{-0.5\height}{\includegraphics[height=0.6in]{branch1-1.png}}
    {\Large \hspace{0.0cm} $\Longrightarrow$ \hspace{0.0cm}}
    \raisebox{-0.5\height}{\includegraphics[height=0.6in]{branch1-1pub.png}}
}

\vspace{0.5\baselineskip}

\item \emph{Lead was~2, others find a block.} The others almost close the gap as the lead drops to~1. The pool publishes its secret blocks, causing everybody to start mining at the head of the previously private branch, since it is longer. The pool obtains a revenue of two. \label{itm:others2}

\vspace{0.5\baselineskip}

\centerline{
    \raisebox{-0.5\height}{\includegraphics[height=0.6in]{branch1-3.png}} 
    {\Large \hspace{0.0cm} $\Longrightarrow$ \hspace{0.0cm}}
    \raisebox{-0.5\height}{\includegraphics[height=0.6in]{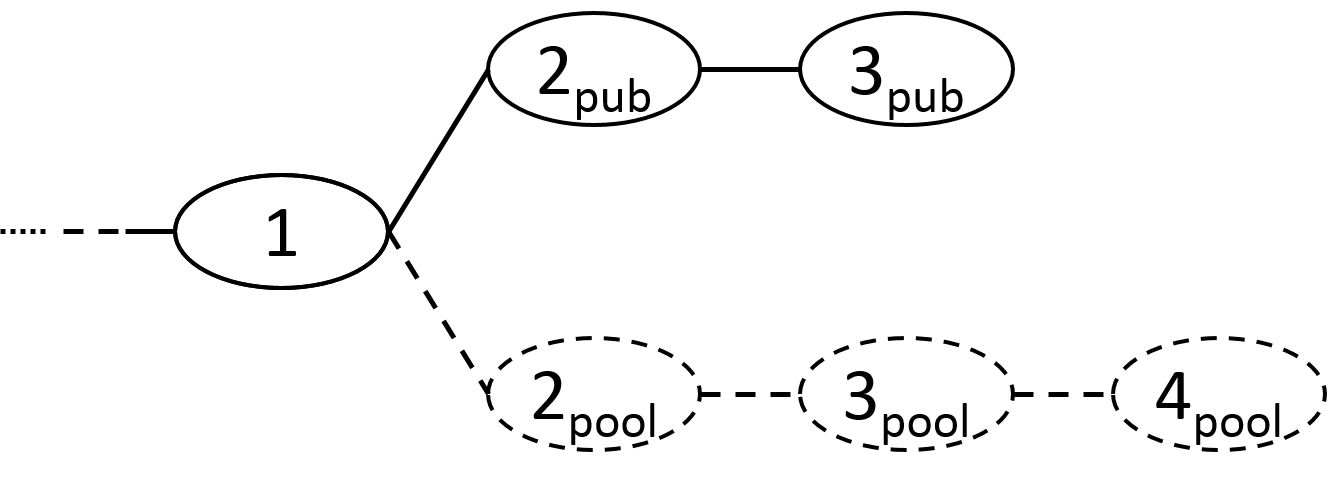}}
    {\Large \hspace{0.0cm} $\Longrightarrow$ \hspace{0.0cm}}
    \raisebox{-0.5\height}{\includegraphics[height=0.6in]{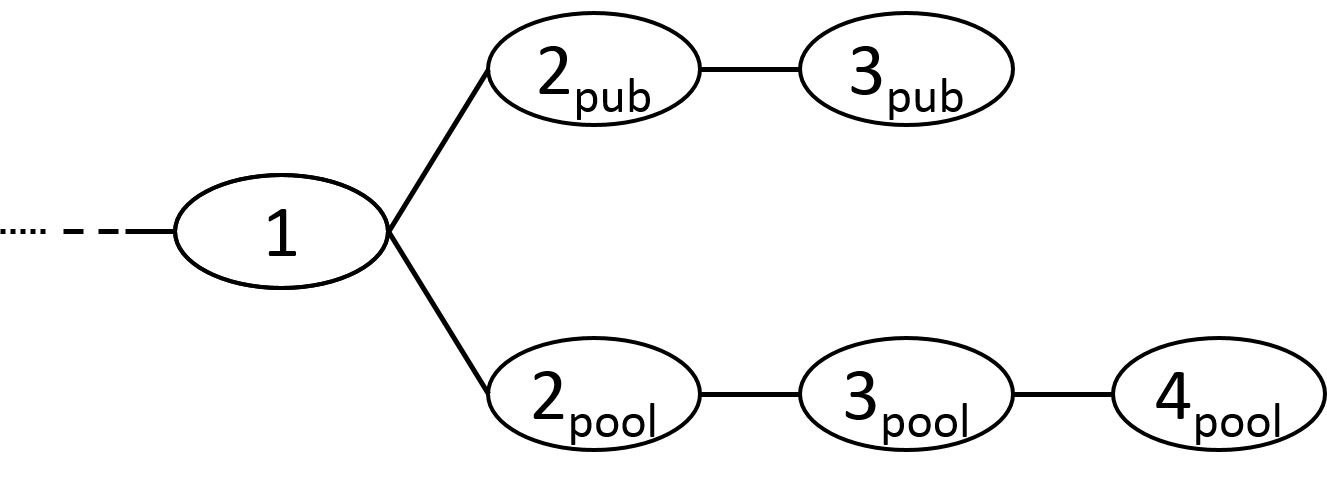}}
}

\vspace{0.5\baselineskip}

\item \emph{Lead was more than 2, others win.} The others decrease the lead, which remains at least two. The new block (say with number $i$) will end outside the chain once the pool publishes its entire branch, therefore the others obtain nothing. However, the pool now reveals its $i$'th block, and obtains a revenue of~one. \label{itm:others3}

\vspace{0.5\baselineskip}

\centerline{
    \raisebox{-0.5\height}{\includegraphics[height=0.6in]{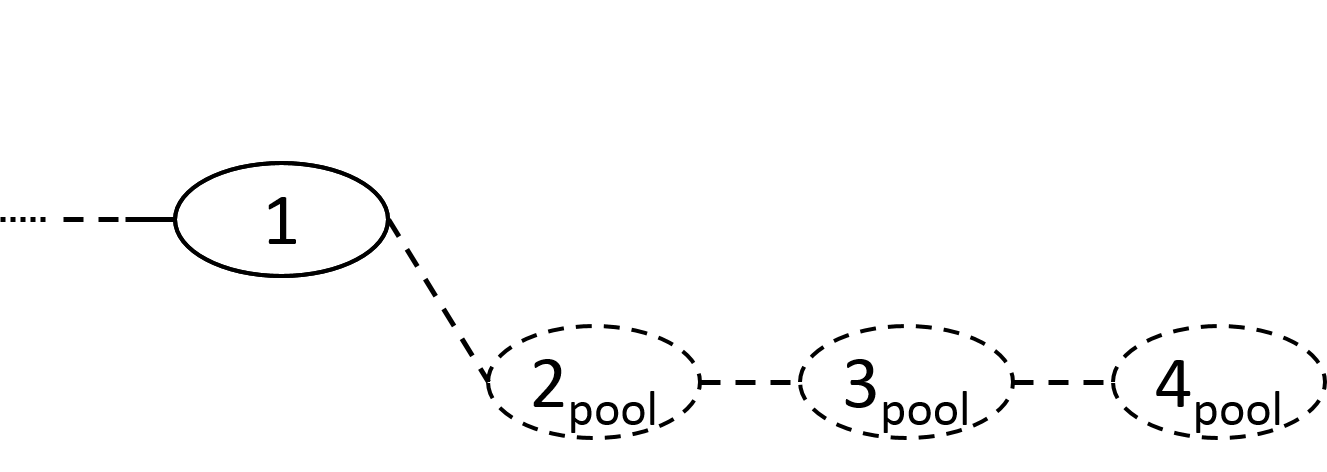}} 
    {\Large \hspace{0.0cm} $\Longrightarrow$ \hspace{0.0cm}}
    \raisebox{-0.5\height}{\includegraphics[height=0.6in]{branch1-3.png}}
    {\Large \hspace{0.0cm} $\Longrightarrow$ \hspace{0.0cm}}
    \raisebox{-0.5\height}{\includegraphics[height=0.6in]{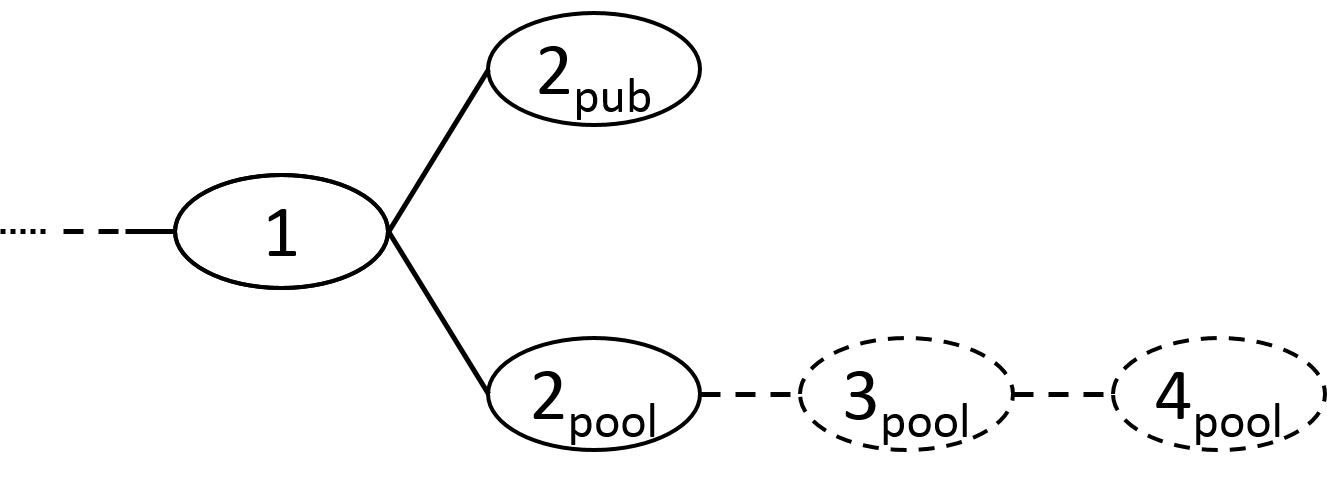}}
}

\vspace{0.5\baselineskip}

\end{enumerate} 

We calculate the revenue of the pool and of the others from the state probabilities and transition frequencies: 
\begin{eqnarray} 
&& r_{\text{others}} = 
        \overbrace{p_{0'} \cdot \gamma (1 - \alpha) \cdot 1}^{\text{Case~\ref{itm:poolOthersZeroPrime}}} + 
        \overbrace{p_{0'} \cdot (1 - \gamma) (1 - \alpha) \cdot 2}^{\text{Case~\ref{itm:othersOthersZeroPrime}}} + 
        \overbrace{p_0 \cdot (1 - \alpha) \cdot 1}^{\text{Case~\ref{itm:others0}}} \label{eqn:rOthers} \\
&& r_{\text{pool}} = 
        \overbrace{p_{0'} \cdot \alpha \cdot 2}^{\text{Case~\ref{itm:poolPoolZeroPrime}}} + 
        \overbrace{p_{0'} \cdot \gamma (1 - \alpha) \cdot 1}^{\text{Case~\ref{itm:poolOthersZeroPrime}}} + 
        \overbrace{p_2 \cdot (1 - \alpha) \cdot 2}^{\text{Case~\ref{itm:others2}}} + 
        \overbrace{P[i > 2] (1 - \alpha) \cdot 1}^{\text{Case~\ref{itm:others3}}} \label{eqn:rPool} \,\,\,\,\,\,\,\,\,\,\,\,\,\,\,\,
\end{eqnarray} 

\newcommand{\RPool}{\ensuremath{ R_{\text{pool}} }}

As expected, the intentional branching brought on by selfish mining leads the honest miners to work on blocks that end up outside the blockchain. This, in turn, leads to a drop in the total block generation rate with $r_{\text{pool}} + r_{\text{others}} < 1$. The protocol will adapt the mining difficulty such that the mining rate at the main chain becomes one block per~10 minutes on average. Therefore, the actual revenue rate of each agent is the \emph{revenue rate ratio}; that is, the ratio of its blocks out of the blocks in the main chain. We substitute the probabilities from (\ref{eqn:pFirst})-(\ref{eqn:pLast}) in the revenue expressions of (\ref{eqn:rOthers})-(\ref{eqn:rPool}) to calculate the pool's revenue for $0 \le \alpha \le \frac{1}{2}$: 
\begin{equation} 
\RPool = 
\frac{r_{\text{pool}}}{r_{\text{pool}} + r_{\text{others}}} = \dots = 
\frac{\alpha (1-\alpha)^2 (4\alpha + \gamma (1 - 2 \alpha)) - \alpha^3}
{1 - \alpha (1 + (2 - \alpha) \alpha)} \,\, . 
\label{eqn:revenue}
\end{equation} 

        \subsection{Simulation} 
        
To validate our theoretical analysis, we compare its result with a Bitcoin protocol simulator. The simulator is constructed to capture all the salient Bitcoin protocol details described in previous sections, except for the cryptopuzzle module that has been replaced by a Monte Carlo simulator that simulates block discovery without actually computing a cryptopuzzle. In this experiment, we use the simulator to simulate~1000 miners mining at identical rates. A subset of $1000\alpha$ miners form a pool running the Selfish-Mine algorithm. The other miners follow the Bitcoin protocol. We assume block propagation time is negligible compared to mining time, as is the case in reality. In the case of two branches of the same length, we artificially divide the non-pool miners such that a ratio of $\gamma$ of them mine on the pool's branch and the rest mine on the other branch. Figure~\ref{fig:revenue} shows that the simulation results match the theoretical analysis. 

        \subsection{The Effect of $\alpha$ and $\gamma$} 

When the pool's revenue given in Equation~\ref{eqn:revenue} is larger than $\alpha$, the pool will earn more than its relative size by using the Selfish-Mine strategy. Its miners will therefore earn more than their relative mining power. Recall that the expression is valid only for $0 \le \alpha \le \frac{1}{2}$. We solve this inequality and phrase the result in the following observation: 

\begin{observation} \label{obs:threshold}
For a given $\gamma$, a pool of size $\alpha$ obtains a revenue larger than its relative size for $\alpha$ in the following range: 
\begin{equation}
\frac{1 - \gamma}{3 - 2\gamma}< \alpha < \frac{1}{2} \,\,\, . 
\label{eqn:threshold}
\end{equation} 
\end{observation}

\begin{figure}[t]
\floatbox[{\capbeside\thisfloatsetup{capbesideposition={right,center},capbesidewidth=0.3\textwidth}}]{figure}[\FBwidth]{
\caption{
Pool revenue using the Selfish-Mine strategy for different propagation factors~$\gamma$, compared to the honest Bitcoin protocol. 
Simulation matches the theoretical analysis, and both show that Selfish-Mine results in higher revenues than the honest protocol above a threshold, which depends on~$\gamma$. 
}
\label{fig:revenue}
}{
\includegraphics[width=0.6\textwidth]{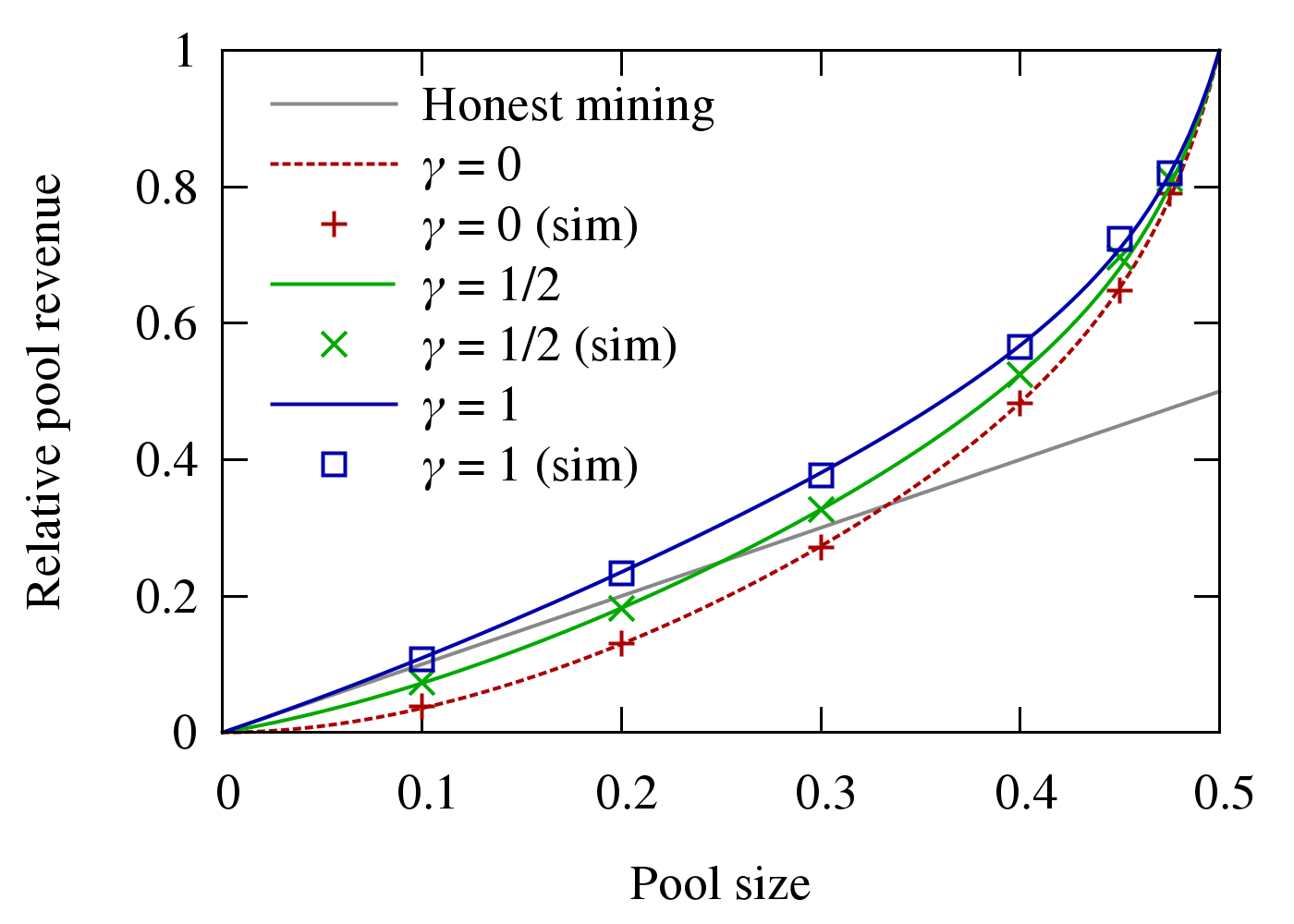}
}
\end{figure}

\begin{figure}[t]
\floatbox[{\capbeside\thisfloatsetup{capbesideposition={right,center},capbesidewidth=0.4\textwidth}}]{figure}[\FBwidth]{
\caption{
For a given $\gamma$, the threshold $\alpha$ shows the minimum power selfish mining pool that will trump the honest protocol. 
The current Bitcoin protocol allows $\gamma=1$, where Selfish-Mine is always superior. 
Unrealistically favorable assumptions leave the threshold at $1/3$. 
We propose (Sec.~\ref{sec:btcPatch}) a protocol change that achieves a threshold of~$1/4$ by setting~$\gamma=0.5$. 
}
\label{fig:threshold}
}{
\includegraphics[width=0.5\textwidth]{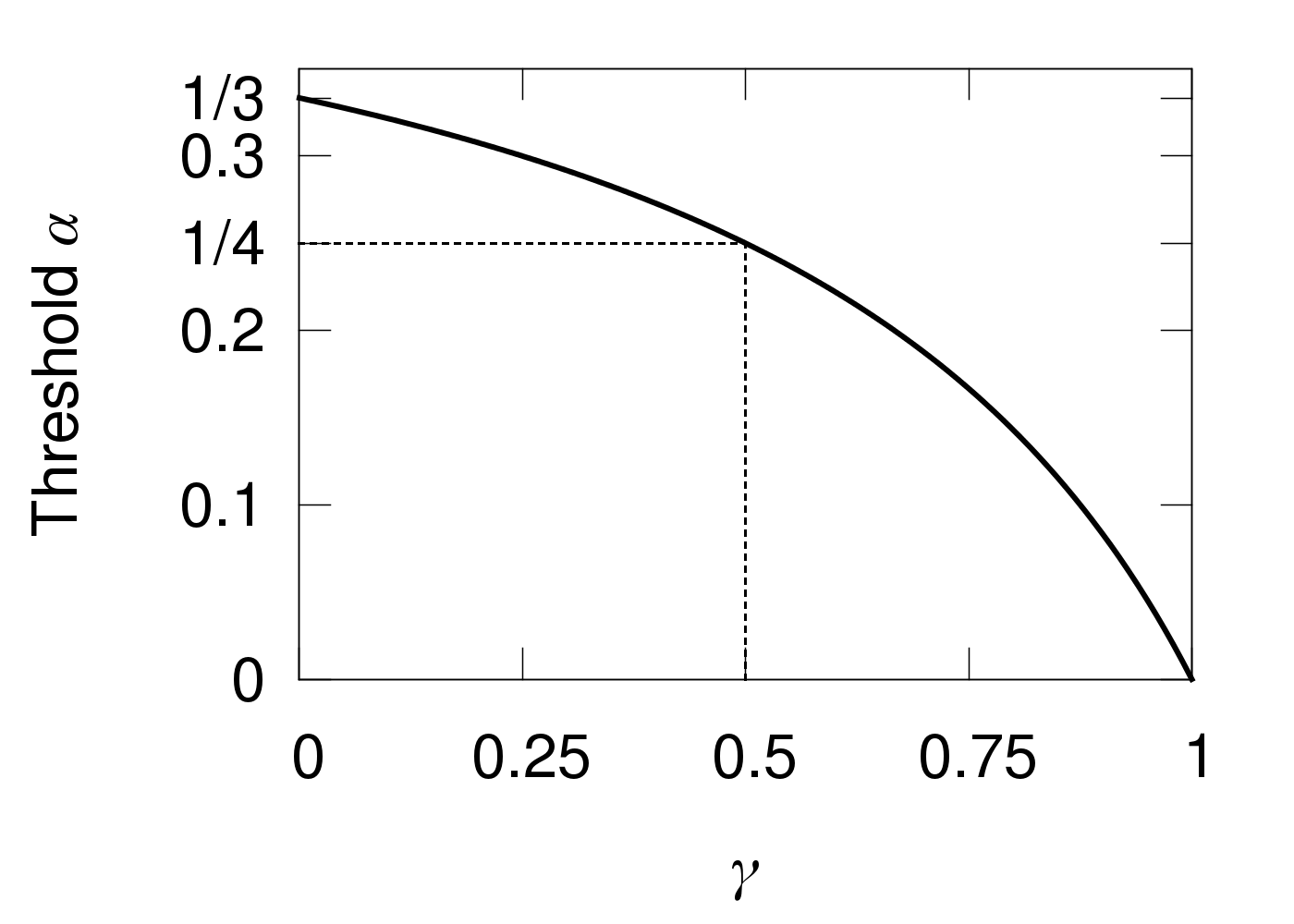}
}
\end{figure}

We illustrate this in Figure~\ref{fig:revenue}, where we see the pool's revenue for different~$\gamma$ values with pool size ranging from~0 (very small pool) to~0.5 (half of the miners). 
Note that the pool is only at risk when it holds exactly one block secret, and the honest miners might publish a block that would compete with it. 
For~$\gamma = 1$, the pool can quickly propagate its one-block branch if the others find their own branch, so all honest miners would still mine on the pool's block. In this case, the pool takes no risk when following the Selfish-Mine strategy and its revenue is always better than when following the correct algorithm. The threshold is therefore zero, and a pool of any size can benefit by following Selfish-Mine. In the other extreme, $\gamma = 0$, the honest miners always publish their block first when the pool has one secret block, and the threshold is at $1/3$. With $\gamma = 1/2$ the threshold is at $1/4$. Figure~\ref{fig:threshold} shows the threshold as a function of $\gamma$. 

We also note that the slope of the pool revenue, \RPool, as a function of the pool size is larger than one above the threshold. This implies the following observation: 

\begin{observation} \label{obs:increasing}
For a pool running the Selfish-Mine strategy, the revenue of each pool member increases with pool size for pools larger than the threshold. 
\end{observation}


    \section{Pool Formation} \label{sec:poolFormation} 

We have shown that once a selfish pool's mining power exceeds the threshold, it can increase its revenue by running Selfish-Mine (Theorem~\ref{obs:threshold}). 
At this point, rational miners will preferentially join the selfish pool to increase their revenues. 
Moreover, the pool's members will want to accept new members, as this would increase their own revenue (Observation~\ref{obs:increasing}). 
The selfish pool would therefore increase in size, unopposed by any mechanism, until it becomes a majority. 
Once a miner pool, selfish or otherwise, reaches a majority, it controls the blockchain. 
The Selfish-Mine strategy then becomes unnecessary, since the others are no longer faster than the pool.
Instead, a majority pool can collect all the system's revenue by following the prescribed Bitcoin protocol, and ignore blocks generated outside the pool; it also has no motivation to accept new members. 
At this point, the currency is not a decentralized currency as originally envisioned. 


    \section{Fixing the Bitcoin Protocol} \label{sec:btcPatch} 
    
Ideally, a robust currency system would be designed to resist attacks by large groups of colluding miners. Since selfish mining attacks yield positive outcomes for group sizes above the threshold, the protocol needs to be amended to set the threshold as high as possible. In this section, we argue that the current Bitcoin protocol has $\gamma \rightarrow 1$, and therefore a threshold of almost zero.
This means a pool of any size can benefit by running Selfish-Mine. We suggest a simple change to the protocol that, if adopted by all non-selfish miners, sets $\gamma$ to $1/2$, and therefore the threshold to $1/4$. This change is backward compatible; that is, any subset of the miners can adopt it without hindering the protocol. Moreover, it is progressive; that is, any ratio of the miners that adopts it decreases $\gamma$, and therefore increases the threshold. 

        \subsection{Problem}
        \negspace

The Bitcoin protocol prescribes that when a miner knows of multiple branches of the same length, it mines and propagates only the first branch it received. 
Recall that a pool that runs the Selfish-Mine strategy and has a lead of~1 publishes its secret block $P$ once it hears of a competing block $X$ found by a non-pool block. If block $P$ reaches a non-pool miner before block $X$, that miner will mine on $P$.  

Because selfish mining is reactive, and it springs into action only after the honest nodes have discovered a block $X$, it may seem to be at a disadvantage. But a savvy pool operator can perform a sybil attack on honest miners by adding a significant number of zero-power miners to the Bitcoin miner network. These virtual miners act as advance sensors by participating in data dissemination, but do not mine new blocks. (Babaioff et al. also acknowledge the feasibility of such a sybil attack~\cite{babaioff2012baloons}). The virtual miners are managed by the pool, and once they hear of block $X$, they ignore it and start propagating block $P$. The random peer-to-peer structure of the Bitcoin overlay network will eventually propagate $X$ to all miners, but the propagation of $X$ under these conditions will be strictly slower than that of block $P$. By adding enough virtual nodes, the pool operator can thus achieve $\gamma$ close to 1. The result, as shown in Equation~\ref{eqn:threshold}, is a threshold close to zero. 

        \subsection{Solution}
        \negspace

We propose a simple, backwards-compatible change to the Bitcoin protocol to address this problem and raise the threshold. 
Specifically, when a miner learns of competing branches of the same length, it should propagate all of them, and choose which one to mine on uniformly at random. 
In the case of two branches of length~1, as discussed in Section~\ref{sec:analysis}, this would result in half the nodes (in expectancy) mining on the pool's branch and the other half mining on the other branch. This yields $\gamma = 1/2$, which in turn yields a threshold of~$1/4$. 

Each miner implementing our change decreases the selfish pool's ability to increase $\gamma$ through control of data propagation. This improvement is independent of the adoption of the change at other miners, therefore it does not require a hard fork. 
This change to the protocol does not introduce new vulnerabilities to the protocol: 
Currently, when there are two branches of equal length, the choice of each miner is arbitrary, effectively determined by the network topology and latency. 
Our change explicitly randomizes this arbitrary choice, and therefore does not introduce new vulnerabilities. 


    \section{Related Work} \label{sec:related} 
    \negspace

Decentralized digital currencies have been proposed before Bitcoin, starting with~\cite{chaum1982blind} and followed by peer-to-peer currencies, e.g.~\cite{vishnumurthy2003karma,yang2003ppay}; see~\cite{miers2013zerocoin,barber2012bitter} for short surveys. 
None of these are centered around a global log; therefore, their techniques and challenges are unrelated to this work. 

Several dozen cryptocurrencies followed Bitcoin's success~\cite{king2012ppcoin,king2013primecoin,wiki2013list}, most prominently Litecoin~\cite{litecoin2013site}. These currencies are based on a global log maintained through mining, and our results apply to all such systems. 

Recent work~\cite{babaioff2012baloons} addresses the lack of incentive for disseminating transactions between miners, since each of them prefers to collect the transaction fee himself. This is unrelated to the mining incentive mechanism we discuss. 

A recent survey~\cite{barber2012bitter} identifies incentive-compatibility as a critical property of the system, and describes a potential scenario that could lead to a single entity controlling a majority of the mining power. 
Our work demonstrates that the Bitcoin system is not incentive compatible, and shows an imminent vulnerability.

A widely cited study~\cite{ron2013quantitative} examines the Bitcoin transaction graph to analyze client behavior. 
The analysis of client behavior is not directly related to our~work. 

The Bitcoin blockchain had one significant bifurcation in March~2013 due to a bug~\cite{gavin2013chainFork}. 
It was solved when the two largest pools at the time manually pruned one branch. 
This bug-induced fork, and the one-off mechanism used to resolve it, are fundamentally different from the intentional forks that Selfish-Mine exploits. 

In a \emph{block withholding attack}, a pool member decreases the pool revenue by never publishing blocks it finds. Although it sounds similar to the strategy of Selfish-Mine, the two are unrelated, as our work that deals with an attack by the pool on the system. 

Various systems build services on top of the Bitcoin global log, e.g., improved coin anonymity~\cite{miers2013zerocoin}, namespace maintenance~\cite{namecoin2013site} and virtual notaries~\cite{sirer2013notary,araoz2013existence}. These services that rely on Bitcoin are at risk in case of a Bitcoin collapse. 


    \negspace
    \section{Discussion} \label{sec:discussion} 
    \negspace

We briefly discuss below several points at the periphery of our scope. 

\paragraph{System Collapse} 

The Bitcoin protocol is designed explicitly to be decentralized. 
We therefore refer to a state in which a single entity controls the entire currency system as a collapse of the Bitcoin system. 

Note that such a collapse does not immediately imply that the value of a Bitcoin drops to~0. 
The controlling entity will have an incentive to accept most transactions, if only to reap their fees, and because if it mines all Bitcoins, it has strong motivation that they maintain their value. 
An analysis of a Bitcoin monopolist's behavior is beyond the scope of this paper, but we believe that a currency controlled by a single entity may deter many of Bitcoin's clients. 

\paragraph{Altruistic Agents} 

Some miners may exhibit altruistic behavior, either following a sense of fairness, community or accomplishment, or to support the system's stability, which is the source of their revenue. Such miners will not participate in Selfish-Mine, whatever the compensation is. 

The presence of such altruistic miners may deter selfish mining attacks if their numbers are sufficient to keep selfish pools below threshold. 
Since Bitcoin mining is open to the public, offering financial rewards, we conjecture that altruistic miners are few, certainly below the hard $2/3$rd bound, and a selfish mining pool of $1/3$ of the system may form. 

\paragraph{Naive Lines of Defense} 

It is easy to hide Selfish-Mine behavior. The selfish pool never identifies itself while mining blocks, and it can publish its blocks~with different IPs and with different Bitcoin addresses at negligible cost. Moreover,~the honest protocol is public, so if a detection mechanism is set up, a selfish pool~knows its parameters and may slightly deviate from the algorithm to avoid them. 

A possible line of defense against selfish mining pools is for counter-attackers to infiltrate selfish pools and to expose their secret blocks for the honest miners. 
However, selfish pool managers can selectively reveal blocks to subsets of the members in the pool, and quickly identify and expel nodes that leak information. 

\paragraph{Imminent danger} 

The Bitcoin system will be immune to Selfish-Mine if there are no pools above the threshold size. Since the threshold is near~0 with the current protocol, this is not the case. 
Miners who value the health of the currency should therefore immediately implement our suggested change to the Bitcoin protocol. 

However, even with our proposed fix that raises the threshold to $25\%$, the outlook is bleak: there already exist pools whose mining power exceeds the $25\%$ threshold~\cite{poolWatch2013poolStats}, and at times, even the $33\%$ theoretical hard limit. 
Responsible miners should therefore break off from large pools until no pool exceeds the threshold size.

These measures must be taken immediately, since if at any time a single pool exceeds threshold it can execute Selfish-Mine, causing a phase transition where rational miners will want to join that pool, leading to a collapse. 


    \section{Conclusion} \label{sec:conclusion}
    \negspace

Bitcoin is the first widely popular cryptocurrency with a broad user base and a rich ecosystem, all hinging on the incentives in place to maintain the critical Bitcoin blockchain.
This paper showed that Bitcoin's mining algorithm is not incentive compatible. The Bitcoin ecosystem is open to manipulation, and potential takeover, by miners seeking to maximize their rewards. This paper
presented Selfish-Mine, a mining strategy that enables pools of colluding miners that adopt it to earn revenues in excess of their mining power. Higher revenues can lead new rational miners to join
selfish miner pools, leading to a collapse of the decentralized currency. This paper showed that the threshold at which selfish mining is effective in the current Bitcoin system is close to zero, and presented
a backwards-compatible modification to Bitcoin that raises this threshold to $1/4$. 

\paragraph{Acknowledgements} 
We are grateful to Raphael Rom, Fred B. Schneider, Eva Tardos, and Dror Kronstein for their valuable advice on drafts of this paper.


\clearpage

\bibliographystyle{splncs} 
\bibliography{btc} 


\clearpage

\appendix

    \section{Probability Calculation} \label{app:probs} 

\begin{figure}[!b] 
\centering
\includegraphics[width=0.6\linewidth]{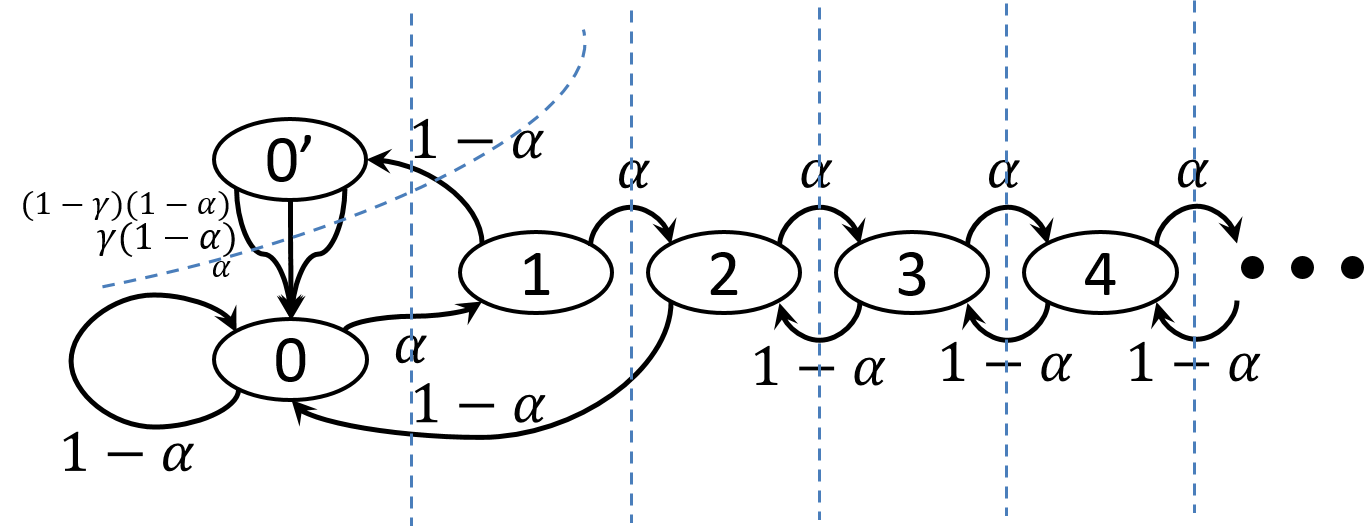}
\caption{ 
State machine with transition frequencies and analysis cuts. 
} 
\label{fig:stateMachineWithCuts} 
\end{figure}

We analyze this state machine to learn the probabilities of it being in its different states. We obtain the following equations with the cuts shown in Figure~\ref{fig:stateMachineWithCuts}: 
\begin{eqnarray} 
&&\alpha p_0 = (1-\alpha) p_1 + (1-\alpha) p_2 \label{eqn:cuts:p0p1p2} \\
&&\alpha p_1 = (1-\alpha) p_2 \label{eqn:cuts:p1p2} \\
&&\forall k \ge 2: \alpha p_k = (1-\alpha) p_{k+1} \label{eqn:cuts:pk} \\
&&p_{0'} = (1 - \alpha) p_1 \label{eqn:cuts:p00p1} \\
&&\sum_{k=0}^\infty p_k + p_{0'} = 1 \label{eqn:cuts:sumP}
\end{eqnarray} 

Equations~\ref{eqn:cuts:p1p2} and~\ref{eqn:cuts:pk} give us 
\begin{equation} 
\forall k \ge 2: p_k = \left( \frac{\alpha}{1-\alpha} \right)^{k-1} p_1 \label{eqn:pk} \,\, .
\end{equation} 

Plugging the expression for $p_2$ from Equation~\ref{eqn:pk} into Equation~\ref{eqn:cuts:p0p1p2} we obtain 
\begin{equation} 
\alpha p_0 = (1-\alpha) p_1 + (1-\alpha) \frac{\alpha}{1-\alpha} p_1 = p_1 \label{eqn:p0p1} \,\, .
\end{equation} 

Now we express all $p$'s in Equation~\ref{eqn:cuts:sumP} as functions of $p_1$ using Equations~\ref{eqn:cuts:p00p1}, \ref{eqn:pk}, and~\ref{eqn:p0p1}: 
\begin{equation*}
1 = p_0 + \sum_{k=1}^\infty p_k + p_{0'} = 
\frac{1}{\alpha} p_1 + \sum_{k=1}^\infty \left( \frac{\alpha}{1-\alpha} \right)^{k-1} p_1 + (1 - \alpha) p_1 \label{eqn:p1Func} \,\, , 
\end{equation*} 
and obtain $p_1$, and therefore the other probabilities: 
\begin{eqnarray}
&& p_1 = \frac{\alpha-2 \alpha^2}{2\alpha^3 - 4\alpha^2 + 1} \\
&& p_0 = \frac{\alpha-2 \alpha^2}{\alpha(2\alpha^3 - 4\alpha^2 + 1)} \\
&& p_{0'} = \frac{(1 - \alpha)(\alpha - 2 \alpha^2)}{1 - 4\alpha^2 + 2\alpha^3} \\
&& \forall k \ge 2: p_k =  \left( \frac{\alpha}{1-\alpha} \right)^{k-1} \frac{\alpha-2 \alpha^2}{2\alpha^3 - 4\alpha^2 + 1}
\end{eqnarray} 


\end{document}